\documentclass[pdflatex,prd,twocolumn,showpacs,superscriptaddress,nofootinbib]{revtex4-1}

\usepackage{bm,amsmath,amssymb}
\usepackage{graphicx}
\usepackage{subfigure}
\usepackage{color}
\usepackage{soul}
\usepackage{hyperref}
\usepackage{multirow}
\usepackage{lineno}
\usepackage{ulem}

\setulcolor{blue}
\setstcolor{red}
\sethlcolor{yellow}

\newcommand \beq {\begin{equation}}
\newcommand \eeq {\end{equation}}
\newcommand \beqa {\begin{eqnarray}}
\newcommand \eeqa {\end{eqnarray}}

\newcommand \hmu {\hat{\mu}}

\def\lsim{\raise0.3ex\hbox{$<$\kern-0.75em\raise-1.1ex\hbox{$\sim$}}}
\def\gsim{\raise0.3ex\hbox{$>$\kern-0.75em\raise-1.1ex\hbox{$\sim$}}}

\begin{document}
\title{Strangeness-Correlations on the pseudo-critical line in (2+1)-flavor QCD}

\author{D. Bollweg}
\affiliation{Computational Science Initiative, Brookhaven National Laboratory, Upton, New York 11973, USA}
\author{H.-T. Ding}
\affiliation{ Key Laboratory of Quark \& Lepton Physics (MOE) and Institute of
Particle Physics, Central China Normal University, Wuhan 430079, China}
\author{J. Goswami}
\affiliation{RIKEN Center for Computational Science,
Kobe 650-0047, Japan}
\author{F. Karsch}
\affiliation{Fakult\"at f\"ur Physik, Universit\"at Bielefeld, D-33615 Bielefeld,
Germany}
\author{Swagato Mukherjee}
\affiliation{Physics Department, Brookhaven National Laboratory, Upton, New York 11973, USA}
\author{P. Petreczky}
\affiliation{Physics Department, Brookhaven National Laboratory, Upton, New York 11973, USA}
\author{C. Schmidt}
\affiliation{Fakult\"at f\"ur Physik, Universit\"at Bielefeld, D-33615 Bielefeld,
Germany}


\begin{abstract}
We present some lattice QCD results on first ($\chi_1^i$) and second ($\chi_2^i$) cumulants of and correlations ($\chi_{11}^{ij}$) among net baryon-number ($B$), strangeness ($S$) and electric charge ($Q$) along
the pseudo-critical line 
($T_{pc}(\mu_B)$) in the temperature ($T$)--baryon chemical potential ($\mu_B$) phase diagram of
(2+1)-flavor QCD. We point out that violations of sum rules among second order
cumulants, which hold in the isospin symmetric limit 
of vanishing electric charge chemical potential, are small 
along the $T_{pc}(\mu_B)$ for
the entire range of 
$\mu_B$
covered in the RHIC beam energy scan. For the 
strangeness neutral matter produced in heavy-ion collisions this leads to a close relation between $\chi_{11}^{BS}$ and $\chi_{11}^{QS}$. We compare lattice QCD results for $\chi_{11}^{BS}/\chi_2^S$
along the $T_{pc}(\mu_B)$ line
with preliminary experimental measurements of  
$\chi_{11}^{BS}/\chi_2^S$ for 
collision energies 
$7.7~{\rm GeV}\le \sqrt{s_{_{NN}}}\le 62.4~{\rm GeV}$. 
While we find good agreements for 
$\sqrt{s_{_{NN}}}\ge 39$~GeV, differences are sizeable at smaller values of $\sqrt{s_{_{NN}}}$. Moreover, we 
compare lattice QCD results for the ratio of the strangeness ($\mu_S$) to baryon ($\mu_B$) chemical potentials, which
define a strangeness neutral system with fixed electric charge to baryon number density, with experimental results obtained by the STAR collaboration for $\mu_S/\mu_B$ using strange baryon yields on the freeze-out line. 
Finally, we determine the baryon chemical potential at the freeze-out ($\mu_B^f$) by comparing $\chi_1^B/\chi_2^B$ along the $T_{pc}(\mu_B)$ with the experimentally measured net-proton cumulants $\chi_1^p/\chi_2^p$. We find that $\{\mu_B^f, T_{pc}(\mu_B^f) \}$ are consistent with the freeze-out parameters of the statistical-model fits to experimentally measured hadron yields for $\sqrt{s_{_{NN}}} \geq 11.5$ GeV.
\end{abstract}


\maketitle


\section{Introduction}

Higher order cumulants of fluctuations of conserved charges and correlations among
different conserved charges of Quantum Chromodynamics (QCD) are sensitive observables
for the occurrence of phase transitions in strongly interacting matter and reflect
the change of relevant degrees of freedom responsible for fluctuations and correlations
in hot and dense strong-interaction matter. In particular, correlations between conserved, net baryon number and
net strangeness number have been 
suggested as sensitive probes \cite{Koch:2005vg} for the 
change of degrees of freedom carrying
strangeness at low temperature (hadrons)
and high temperature (quarks), respectively.
The temperature and density dependence
of conserved charge fluctuations and 
their higher order cumulants
is studied in lattice 
QCD calculations using Taylor series 
expansions as well as in numerical
simulations with imaginary chemical
potentials \cite{Gavai:2001fr,Allton:2002zi,DElia:2002tig,Borsanyi:2011sw}. 

Experimentally the cumulants
of conserved charge densities are not
directly accessible. E.g.
net proton \cite{STAR:2020tga}
or kaon number \cite{NA49:2002pzu,ALICE:2013mez,STAR:2019bjj} fluctuations
and correlations among them
are used as proxies for 
baryon or strangeness number fluctuations. Incorporating
strange baryon contributions also allows
to construct proxies for strangeness and
baryon number as well as strangeness 
and electric charge correlations \cite{Braun-Munzinger:2014lba,STAR:2019bjj,Bellwied:2019pxh}.

The experimentally observed charge
fluctuations and correlations are expected to be generated
at the so-called chemical freeze-out
temperature $T_{f}$. Also this temperature
and the baryon as well as strangeness
chemical potentials ($\mu_B^f,\mu_S^f)$ that control the 
thermal conditions at the time of freeze-out are not directly accessible
in heavy ion collisions.
Eventually they could be deduced through
a comparison of experimental data with
QCD thermodynamics, if the medium is 
in equilibrium at the time of freeze-out.
In practice the experimental determination of freeze-out parameters
involves a model dependent step. Experimentally determined
particle yields are compared to hadronization models
\cite{Andronic:2017pug,Adamczyk:2017iwn,ALICE:2023ulv}. The thus determined freeze-out
temperature at various values of the 
beam energy, $T_f(\sqrt{s_{_{NN}}})$,
is found to be in 
good agreement with the pseudo-critical
temperature, $T_{pc}(\hmu_B)$, that is determined in lattice QCD calculations
\cite{Bazavov:2018mes,Borsanyi:2020fev}.
Generally this is taken as support for 
the assumption that hadron resonance gas
(HRG) models 
provide a good approximation to the 
thermodynamics of strong-interaction matter at the time of freeze-out.

Aside from studies of higher order
cumulants of proxies for net baryon-number fluctuations, the calculation of 
correlations between low order cumulants of conserved charge densities, e.g.
correlation of net strangeness and net
baryon-number, is being performed in the beam energy scan at 
RHIC. We previously presented comparisons of higher order cumulants of net baryon-number fluctuations
with results obtained by the STAR collaboration \cite{Bazavov:2020bjn,HotQCD:2017qwq} 
for the related proton number cumulants.
We also compared results of the STAR collaboration on strange hadron
yields at several values of the beam energy \cite{STAR:2019bjj} with lattice QCD results \cite{Bazavov:2014xya,Bollweg:2020pjb} 
on correlations
between net baryon-number and strangeness. 
This provided evidence for the influence
of additional strange hadrons, not listed by the Particle Data Group as established
strange hadron resonances \cite{ParticleDataGroup:2022pth}. 

We previously also calculated low order
cumulants of net baryon-number fluctuations at
non-zero values of the baryon chemical potential,
using high statistics datasets generated in (2+1)-flavor QCD calculations at finite temperature
\cite{Bollweg:2022rps,Bollweg:2021vqf}. In these
calculations we focused on an analysis
of second order cumulants at vanishing
values of the chemical potentials as
well as an analysis 
of the convergence of the Taylor series 
and the comparison with Pad\'e resummed
results obtained from such series expansions. We concluded that expansions
of second order cumulants, in particular
the second order cumulant for net baryon-number fluctuations, are well controlled
for baryon chemical potentials $\mu_B/T\lesssim 1.5$.

We extend this analysis here by calculating correlations between 
conserved charge densities as function
of the baryon chemical potential on the
pseudo-critical line in (2+1)-flavor QCD
using Taylor series for second order
cumulants up to NNLO order, {\it i.e.}
up to ${\cal O}(\mu_B^4)$. We will use
these results to compare with recent 
preliminary results of the STAR collaboration obtained for correlations between
net baryon number and strangeness densities \cite{STAR-CPOD}.

The data sets used for our calculation and further details relevant 
for our data analysis, have been described
previously in section II of  \cite{Bazavov:2020bjn} 
and the number of 
configurations, used for our current work, are given in
Table I of \cite{Bazavov:2020bjn}. In particular, as
discussed in \cite{Bazavov:2020bjn},
using fits of lattice QCD results obtained at different values of the
lattice spacing to rational polynomials, 
we obtain continuum extrapolated estimates for the 
$\hmu_B$-dependence of various second
order cumulants on the pseudo-critical line,
$T_{pc}(\hmu_B)$, as well as extrapolations performed directly for ratios of second order cumulants.

This paper is organized as follows.
In Section II we introduce basic 
formulae used for the calculation of Taylor expansions of second order cumulants. In Section III we present 
results for various second order cumulants obtained in lattice QCD
calculations as functions of $T$ and $\hmu_B$. In Section IV we present
results for these cumulants 
evaluated on the pseudo-critical line $T_{pc}(\mu_B)$ and compare them with
recent results of the STAR collaboration obtained during the 
beam energy runs BES-II at RHIC.
We give our conclusions in Section V. In an Appendix we give some explicit expressions for the expansion
of second order cumulants up to ${\cal O}(\mu_B^4)$.

\section{Cumulants of conserved charge densities}
\label{sec:cumulants}

Cumulants
of conserved charge fluctuations at vanishing chemical
potentials for the conserved charges of $(2+1)$-flavor QCD,
{\it i.e.} net baryon-number ($B$),
electric charge ($Q$) and strangeness ($S$),
are obtained from the QCD partition function,
$\mathcal{Z}(T,V,\vec{\mu})$, as derivatives with
respect to the associated chemical potentials
$\vec{\mu}=(\mu_B, \mu_Q, \mu_S)$,
\begin{equation}
\chi_{ijk}^{BQS}(T,V) =\left. 
\frac{1}{VT^3}\frac{\partial \ln\mathcal{Z}(T,V,\vec{\mu}) }{\partial\hmu_B^i \partial\hmu_Q^j \partial\hmu_S^k}\right|_{\vec{\mu}=0} \; ,
\label{suscept}
\end{equation}
with $\hat{\mu}_X\equiv \mu_X/T$.
These cumulants can be used to set up
Taylor series for cumulants at non-zero
values of the chemical potentials. 
Starting with the Taylor series for the pressure,
\begin{eqnarray}
        \frac{P}{T^4} &=& \frac{1}{VT^3}\ln\mathcal{Z}(T,V,\vec{\mu}) 
        = \sum_{i,j,k=0}^\infty
\frac{\chi_{ijk}^{BQS}}{i!j!\,k!} \hmu_B^i \hmu_Q^j \hmu_S^k \; ,
\label{Pdefinition}
\end{eqnarray}
we easily obtain series for second order 
cumulants by taking appropriate derivatives with respect to the chemical potentials. In these Taylor series 
we replace the chemical
potentials $\hmu_Q$ and $\hmu_S$ by series
in terms of $\hmu_B$ only, 
\begin{eqnarray}
\mu_S/\mu_B &=& \sum_{i=0}^\infty s_{2 i+1} (T)\ \hmu_B^{2 i+1} \; , \label{muSexp}\\
\mu_Q/\mu_B &=& \sum_{i=0}^\infty q_{2 i+1} (T)\ \hmu_B^{2 i+1} \; ,
\label{muQexp}
\end{eqnarray}
where the coefficients $s_i$, $q_i$ are
fixed through constraints on the net baryon-number density\footnote{We use here and in the following dimensionless number densities. I.e. densities $n_X$, $X=B,\ Q,\ S$, are given in units of $T^3$.}, $n_B$, and the 
ratio of electric charge and strangeness
densities, $n_Q/n_S$. 
We demand strangeness neutrality, $n_S=0$, and a fixed ratio of net electric charge and baryon number densities,
$n_Q/n_B=0.4$, which is consistent with experimental conditions realized in heavy ion collisions of lead-lead or gold-gold nuclei. 
The resultant 
coefficients $s_i$, $q_i$ are given in
\cite{Bazavov:2017dus} for $i=1,3,5$.
With this we obtain for the next-to-leading order (NLO) expansion of
second order cumulants
\begin{eqnarray}
\chi_{11}^{BS}(T,\vec{\mu})&=& \chi_{11}^{BS} + \frac{\hmu_B^2}{2!}\big[\chi_{31}^{BS} + 2s_1\chi_{22}^{BS} + s_1^2\chi_{13}^{BS} \label{BSNLO}\\
&&\hspace{-0.4cm}+ 2 q_1\chi_{211}^{BQS} + 2 q_1s_1 \chi_{112}^{BQS} + q_1^2 \chi_{121}^{BQS}\big] + {\cal O}(\hmu_B^4) \; ,
\nonumber 
\\
\chi_{11}^{QS}(T,\vec{\mu})&=&\chi_{11}^{QS}+\frac{\hmu_B^2}{2!}\big[\chi_{211}^{BQS} + 2s_1\chi_{112}^{BQS} + s_1^2\chi_{13}^{QS} \label{QSNLO}\\
&&\hspace{-0.4cm}+  2 q_1\chi_{121}^{BQS} + 2 q_1s_1 \chi_{22}^{QS} + q_1^2 \chi_{31}^{QS}\big] + {\cal O}(\hmu_B^4)\; , 
\nonumber 
\\
\chi_{2}^{S}(T,\vec{\mu})&=& \chi_{2}^S +\frac{\hmu_B^2}{2!} \big[\chi_{22}^{BS} + s_1^2\chi_{4}^S + 2 s_1\chi_{13}^{BS} \label{SNLO}\\
&&\hspace{-0.4cm}+ 2q_1\chi_{112}^{BQS}+2q_1s_1\chi_{13}^{QS}+ q_1^2\chi_{22}^{QS}\big]+{\cal O}(\hmu_B^4) \; ,
\nonumber
\\
\chi_{2}^{B}(T,\vec{\mu})&=& \chi_{2}^B +\frac{\hmu_B^2}{2!} \big[ \chi_4^B+s_1^2 \chi_{22}^{BS}+2 s_1\chi_{31}^{BS} \label{BNLO}\\ 
&&\hspace{-0.4cm}+ 2 q_1\chi_{31}^{BQ}+2 q_1 s_1 \chi_{211}^{BQS}+q_1^2\chi_{22}^{BQ} \big]+{\cal O}(\hmu_B^4) \; .
\nonumber
\end{eqnarray}
Explicit expressions for the order
${\cal O}(\hmu_B^4)$  (NNLO) expansion coefficients, which we will use in the 
following, are given in Appendix A.
The leading order $\mu_S/\mu_B$ is related with the $\chi_{11}^{BS}/\chi_2^S$ in the following way,
\begin{equation}
    \frac{\mu_S}{\mu_B} = -\frac{\chi_{11}^{BS}(T,0)}{\chi_2^S(T,0)}-\frac{\chi_{11}^{QS}(T,0)}{\chi_2^S(T,0)}q_1 + \mathcal{O}(\mu_B^2)\,.
\end{equation}

\begin{figure*}[t]
\includegraphics[scale=0.54]{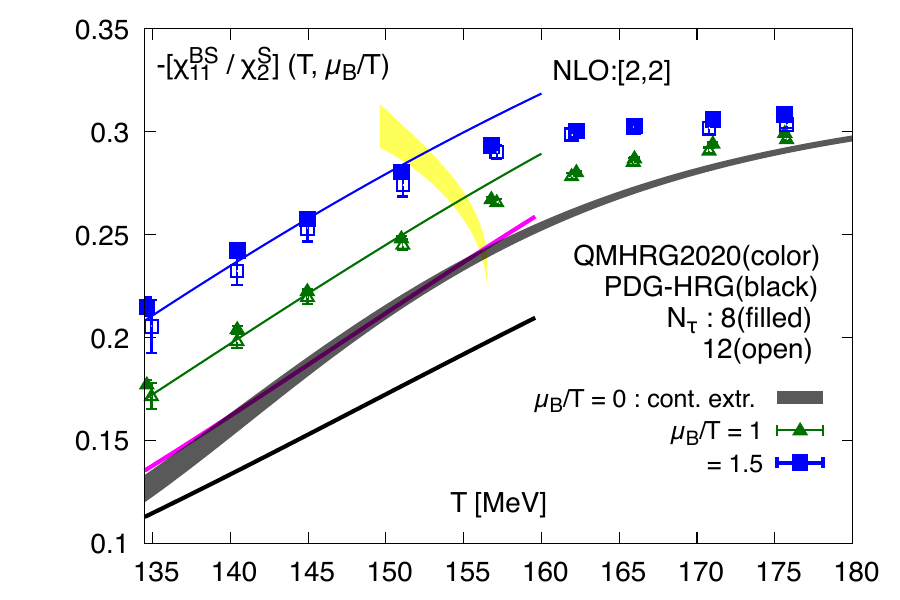}
\includegraphics[scale=0.54]{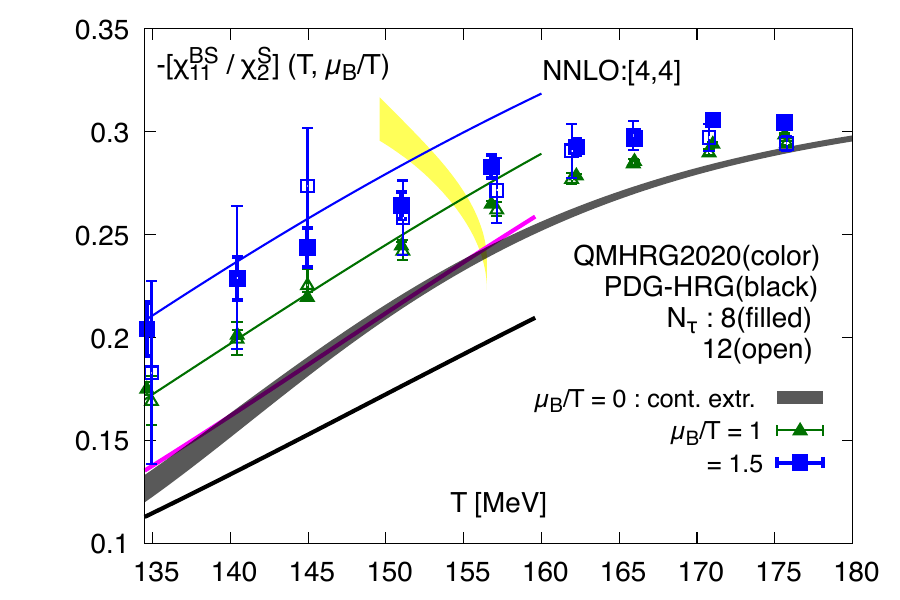}
\includegraphics[scale=0.54]{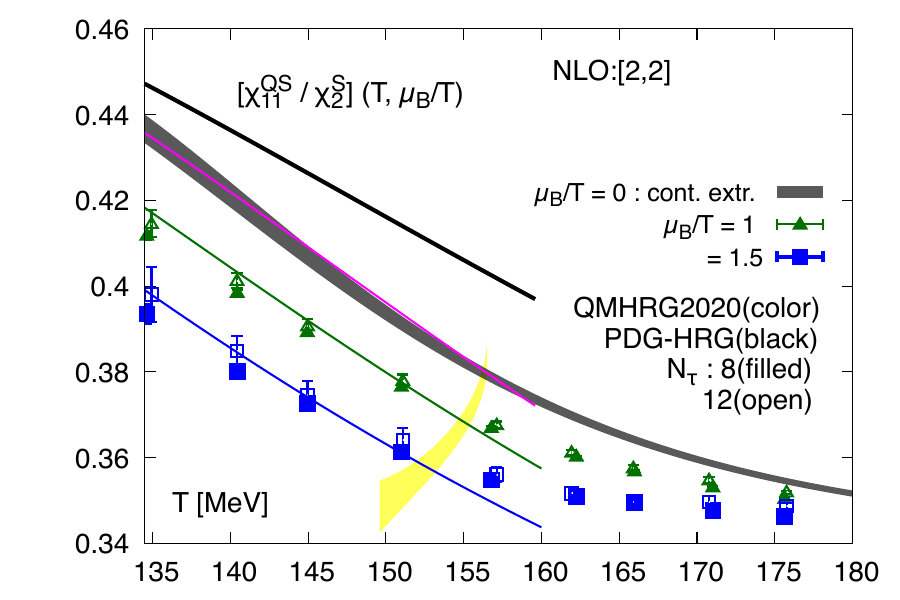}
\includegraphics[scale=0.54]{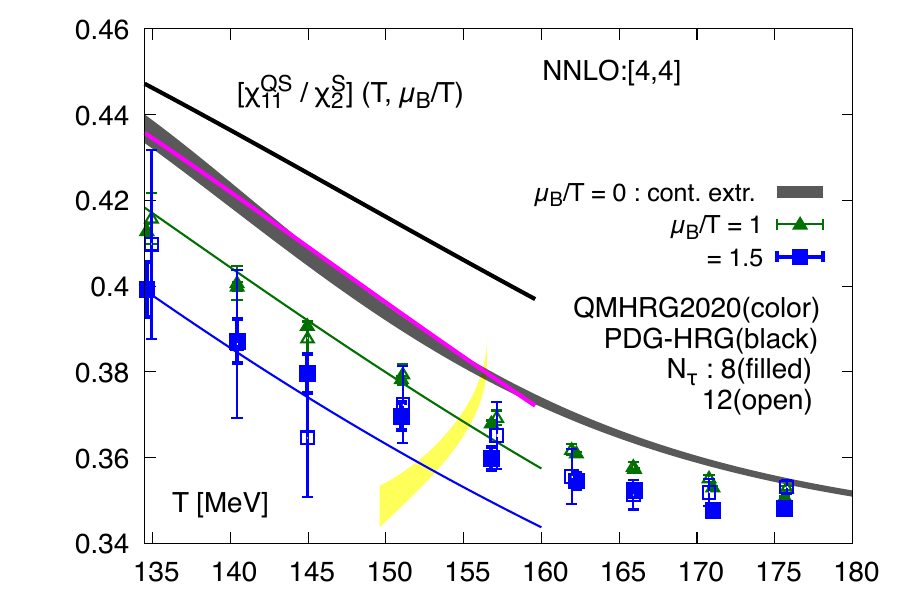}
\caption{{\it Top:} Correlation between net baryon-number and strangeness number densities normalized to the second order cumulant of strangeness fluctuations versus $T$
for several values of $\hmu_B$. Shown 
are results obtained on lattices with temporal extent $N_\tau=8$ (filled symbols) and $12$ (open symbols) from ${\cal O}(\hmu_B^2)$ (left) and ${\cal O}(\hmu_B^4)$ (right) Taylor series for $\chi_{11}^{BS}$ and $\chi_2^S$, respectively. The lattice QCD results
are compared to HRG model calculations
using the QMHRG2020 list (colored lines). For $\hmu_B=0$
also a result based on the PDG-HRG list
is shown (black line). 
The grey band shows the continuum extrapolated results for the second order cumulant ratio at
vanishing chemical potential \cite{Bollweg:2021vqf}.
The yellow band shows the location of the pseudo-critical line $T_{pc}(\hmu_B)$.  
{\it Bottom:}
same as figures on the top, but for correlation between net electric charge and strangeness densities normalized to the second order cumulant of strangeness fluctuations. 
}
\label{fig:BSmudep}
\end{figure*}

In the case of degenerate up and down quark masses, which generally is used in lattice QCD calculations, the cumulants  $\chi_{11}^{BS}$, $\chi_{11}^{QS}$, and $\chi_{2}^{S}$ 
are directly related to each other,
\begin{equation}
2 \frac{\chi_{11}^{QS}(T,\vec{\mu})}{\chi_{2}^S(T,\vec{\mu})} - 
\frac{\chi_{11}^{BS}(T,\vec{\mu})}{\chi_{2}^S(T,\vec{\mu})}= 1 + \frac{\Delta^{BQS}(T,\vec{\mu})}{\chi_{2}^S(T,\vec{\mu})} 
\; .
\label{BS-QS-rel}
\end{equation}
For $\mu_Q\ne 0$, one finds,
\begin{eqnarray}
\Delta^{BQS}(T,\vec{\mu})&&= q_1 \frac{\hmu_B^2}{2}
\Bigg( q_1(2  \chi^{BQS}_{031}
-\chi^{BQS}_{022} -\chi^{BQS}_{121})
\nonumber\\
&&-s_1 (2 \chi^{BQS}_{013}-4
    \chi^{BQS}_{022}+ 2 \chi^{BQS}_{112}
    ) \nonumber \\
&&- \chi^{BQS}_{103}+4 \chi^{BQS}_{121}-2
    \chi^{BQS}_{202}-\chi^{BQS}_{301}
\Bigg)\nonumber\\
&&+{\cal O}(\hmu_B^4)\; .
\label{DeltaBQS}
\end{eqnarray}

At vanishing electric charge chemical potential $\mu_Q$, {\it i.e.} in the isospin symmetric case with $n_Q/n_B=0.5$,
one thus obtains the sum rule,
\begin{eqnarray}
\Delta^{BQS}(T,\vec{\mu})&=& 0 
\;\; ,\;{\rm for} \; \vec{\mu}=(\mu_B,0,\mu_S)
\; .
\label{2QS-BS-S2}
\end{eqnarray}

We will compare results for the second order
cumulants with HRG model calculations
based on the list of hadrons published in the tables of the Particle Data group \cite{ParticleDataGroup:2022pth} (PDG-HRG) as well as the QMHRG2020 list compiled by the HotQCD
Collaboration\cite{Bollweg:2021vqf,bollweg2021dataset}. In order to probe the validity
of Eq.~\ref{2QS-BS-S2} and the isospin
symmetry violations induced by a non-vanishing
electric charge chemical potential in HRG
model calculations, we introduced an 
isospin symmetric version of these hadron lists. For this purpose we set masses for 
charged strange hadrons equal to the masses of their neutral
partners. This concerns 7 strange hadrons, for which
mass differences are listed in the PDG tables:
$K, K^{*}, K_2^{*},\Sigma,\Sigma(1385),\Xi,\Xi(1530)$. In all these cases
isospin violations, arising from the mass differences between charged and neutral strange hadrons, are below 1\%. We note that
setting the masses equal leads only to small changes in 
the Boltzmann weights entering HRG model calculations. In
the vicinity of the pseudo-critical temperature $T_{pc,0}= 156.5(1.5)$~MeV
\cite{Bazavov:2018mes} this amounts to
about $0.6$\%. In general this small
difference is negligible. However, when comparing QCD and HRG model results for 
$\Delta^{BQS}$ as
given in Eq.~\ref{DeltaBQS}, this small
change leads to differences which are 
of the same magnitude as the modifications induced by a non-vanishing
electric charge chemical potential itself.

\section{Second order cumulants as a function
of temperature and baryon chemical potential}

In \cite{Bollweg:2021vqf} we presented 
continuum extrapolated results for 
all second order cumulants at vanishing
values of the chemical potentials.
For non-vanishing values of the chemical
potentials Taylor series for the diagonal
second order net baryon-number cumulant,
$\chi_2^B(T,\hmu_B)$,
have been presented in \cite{Bazavov:2020bjn}.
Some preliminary results on 
strangeness correlations and fluctuations at non-zero $\hmu_B$ have also been shown in \cite{Bollweg:2020pjb,Karsch:2017zzw}.

Here we focus on the behavior of off-diagonal, second order cumulants. We concentrate on a discussion
of correlations between net strangeness number
($n_S$) and  net baryon-number ($n_B$) densities as well as  $n_S$ and
net electric charge ($n_Q$) density.
We shall consider thermal conditions that 
are of relevance for comparisons with
experimental conditions realized in heavy ion collisions, {\it i.e.} we consider 
a thermal medium
with vanishing net strangeness, $n_S=0$, and $n_Q/n_B=0.4$. 

In Fig.~\ref{fig:BSmudep} we show
the ratios $-\chi_{11}^{BS}(T,\hmu_B)/\chi_2^S(T,\hmu_B)$ (top) and 
$\chi_{11}^{QS}(T,\hmu_B)/\chi_2^S(T,\hmu_B)$ (bottom),
obtained from ${\cal O}(\hmu_B^2)$ (left) and ${\cal O}(\hmu_B^4)$ (right) Taylor series,
respectively. 
As can be seen these expansions 
agree well for small $\hmu_B$. The 
second and fourth order expansions start
to differ at temperatures below the 
pseudo-critical line for the chiral transition (yellow bands) at $\hmu_B\simeq 1.5$.

The lattice QCD results
are compared to HRG model calculations
that utilize the PDG-HRG and QMHRG2020
lists. As noted already in the 
analysis of second order cumulants at
vanishing chemical potential \cite{Bollweg:2021vqf} HRG model
calculations based on the QMHRG2020 list describe the lattice QCD results
well at low temperatures up to $T_{pc}(\hmu_B)$. HRG model calculations based on the PDG-HRG list
\cite{ParticleDataGroup:2022pth}, however, are insufficient 
already for $T\gsim 130$~MeV.
While QMHRG2020 and lattice QCD 
calculations agree well at $T_{pc,0}\equiv T_{pc}(0)$,
differences become larger with increasing
$\hmu_B$ also on the pseudo-critical line.

The pseudo-critical line
has been determined
in lattice QCD calculations as a Taylor 
series up
to ${\cal O}(\hmu_B^4)$
\cite{Bonati:2015bha,Bazavov:2018mes,Borsanyi:2020fev},
\begin{equation}
T_{pc}(\mu_B) = T_{pc,0}\left[ 1 - \kappa_2 \hmu_B^2 +
\kappa_4 \hmu_B^4 \right] \; .
  \label{Tpcmu}
  \end{equation}
We use results from \cite{Bazavov:2018mes},
{\it i.e.} $T_{pc,0}=(156.5\pm 1.5)$~MeV and $\kappa_2=0.012(4)$. Within errors this is 
in agreement with results obtained in \cite{Bonati:2015bha,Borsanyi:2020fev}.
The ${\cal O}(\hmu_B^4)$ correction in
Eq.~\ref{Tpcmu}
has been found to vanish within current
statistical errors \cite{Bazavov:2018mes,Borsanyi:2020fev}. We leave out the contribution of the $O(\hat\mu_B^4)$ in the following
as it as been found to vanish within current statistical errors \cite{Bazavov:2018mes,Borsanyi:2020fev}.
I.e. Ref. \cite{Bazavov:2018mes} found $\kappa_4=0.000(4)$ and Ref. \cite{Borsanyi:2020fev} reported 
$\kappa_4=0.00032(67)$. The pseudo-critical temperature 
$T_{pc}(\hmu_B)$
is shown in Fig.~\ref{fig:BSmudep}
as a  yellow band.

As can be seen in Fig.~\ref{fig:BSmudep}
the magnitude of the ratio $\chi_{11}^{BS}(T,\hmu_B)/\chi_2^S(T,\hmu_B)$ increases with increasing $\hmu_B$.
On the pseudo-critical line this amounts
to only a moderate change of about (15-20)\% between $\hmu_B=0$ and 1.5. 

\begin{figure}[t]
\includegraphics[width=0.45\textwidth]{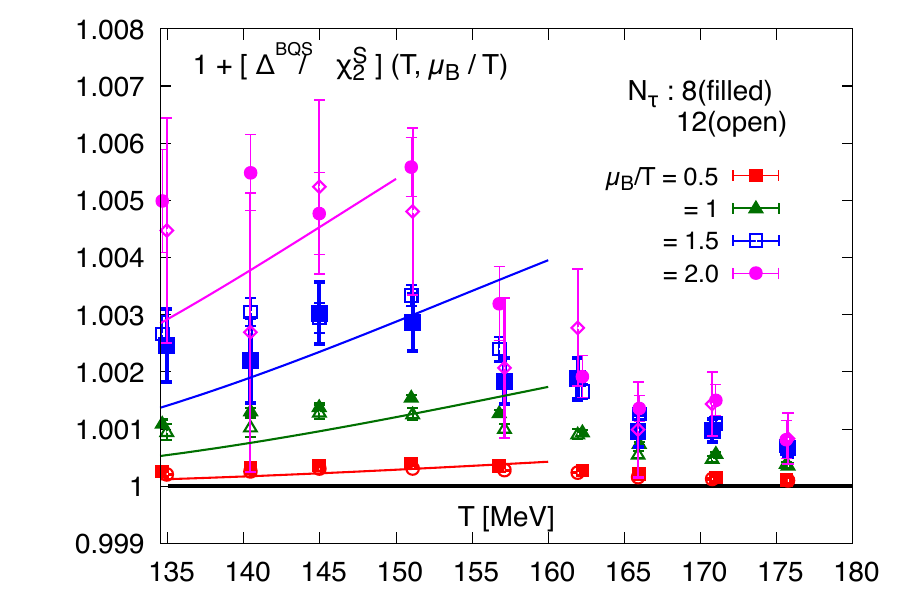}
\caption{Deviations from the sum rule,
Eq.~\ref{BS-QS-sumrule},
valid in isospin symmetric matter for the relation between 
strangeness and baryon-number correlation
on the one hand, and strangeness and 
electric charge correlations on the other hand. Shown are lattice QCD results using 
$4^{\rm th}$ order Taylor series results for the second order cumulants. Also shown are HRG model calculations using the QMHRG2020 list 
with additional isospin symmetrization for 7 strange hadrons as discussed in Sec.~\ref{sec:cumulants}.
}
\label{fig:QS-BS-S2-relation}
\end{figure}

In the previous section we pointed out that the sum rule, $2\chi_{11}^{QS}-\chi_{11}^{BS}=\chi_2^S$,
holds in QCD at any value of the temperature for isospin symmetric matter,
{\it i.e.} for $\mu_Q=0$. 
The difference of the cumulant ratios $\chi_{11}^{BS}/\chi_2^S$ and
$2\chi_{11}^{QS}/\chi_2^S$ thus equals unity 
for all $\hmu_Q=0$,
\begin{equation}
 2 \frac{\chi_{11}^{QS}(T,\vec{\mu})}{\chi_{2}^S(T,\vec{\mu})} - 
\frac{\chi_{11}^{BS}(T,\vec{\mu})}{\chi_{2}^S(T,\vec{\mu})}= 1 \;\; {\rm for}\;\; \vec{\mu}=(\mu_B,0,\mu_S) \\
\; .
\label{BS-QS-sumrule}   
\end{equation}
For $\hmu_Q\ne 0$, e.g. for $n_Q/n_B=0.4$,
we find that deviations from unity 
increase with increasing $\hmu_B$. 
Deviations are largest in the vicinity
of $T_{pc}(\hmu_B)$. However, they stay
below\footnote{Note that isospin
violations in QCD at non-zero temperature are indeed expected to
be negligible \cite{Pisarski:1983ms}. Nonetheless large violations of isospin symmetry have been reported
recently to occur in kaon production in high-energy nucleus-nucleus collisions \cite{Brylinski:2023nrb,NA61SHINE:2023azp}.}
0.5\% even for $\hmu_B\simeq 2$.
This is evident from the fourth order
Taylor series results shown in 
Fig.~\ref{fig:QS-BS-S2-relation}. 
At $\hmu_B=2$ the
pseudo-critical temperature is about
$T_{pc}(\hmu_B=2)\simeq 150$~MeV. 
The corresponding baryon chemical
potential thus is $\mu_B\simeq 300$~MeV,
which corresponds to conditions reached 
in heavy ion collisions 
at beam energies below $11.5$~GeV. We thus expect
Eq.~\ref{BS-QS-sumrule} to hold for
correlations of net baryon-number or electric charge with
net strangeness number 
at the LHC and also at RHIC for almost the entire
range of beam energies covered in the beam energy scan.

\begin{figure}[t]
\includegraphics[scale=0.52]{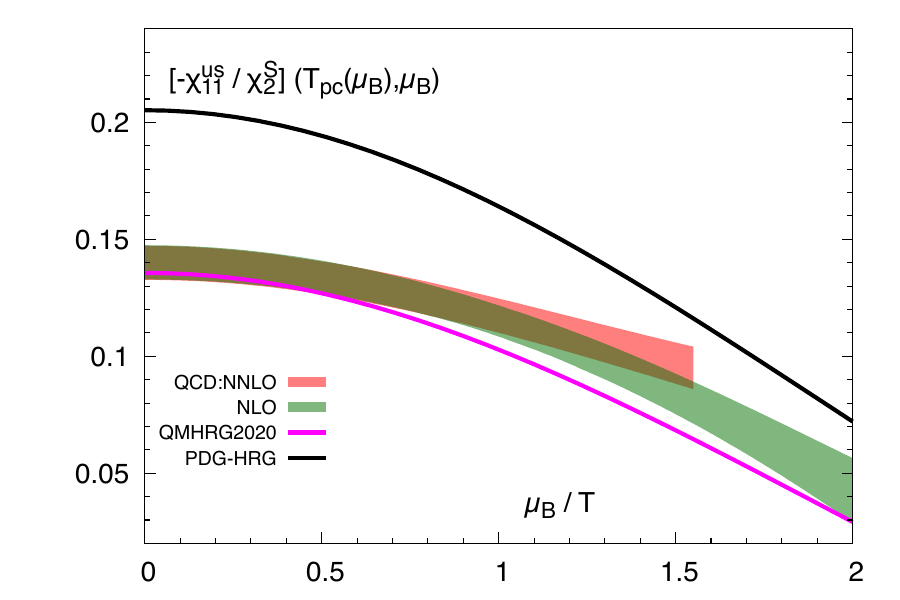}
\caption{Correlation between light and strange
quark-number densities on the pseudo-critical line. Shown are continuum estimates for second and fourth 
order Taylor expansion results for 
$-\chi_{11}^{us}/\chi_2^S$ on the pseudo-critical line 
that have been obtained from fits 
to data obtained in simulations
on lattices with temporal extent $N_\tau=8$ and 12. These results 
are compared with
HRG model results based on the PDG-HRG
and QMHRG2020 hadron lists, respectively.
}
\label{fig:BS-QS-mu_pseudo}
\end{figure}

We finally also note that the sum of $\chi_{11}^{BS}$ and $\chi_{11}^{QS}$
gives the correlation between light 
and strange quark densities for 
any value of the chemical potentials $\vec{\mu}$,  
\begin{equation}
\chi_{11}^{us}(T,\vec{\mu}) = - \left( \chi_{11}^{BS}(T,\vec{\mu}) + \chi_{11}^{QS}(T,\vec{\mu}) \right) \; .
\label{us}
\end{equation}
Making use of the very well satisfied
sum rule between the second order strange cumulants (cf. Eq.~\ref{BS-QS-sumrule}) we 
thus find that $\chi_{11}^{BS}$ also
provides a good approximation for the
correlation between
light and strange quark densities,
\begin{eqnarray}
    \frac{\chi_{11}^{us}(T,\hmu_B)}{\chi_2^S(T,\hmu_B)} 
    &=& -\left( \frac{\chi_{11}^{BS}(T,\hmu_B)}{\chi_2^S(T,\hmu_B)} +\frac{\chi_{11}^{QS}(T,\hmu_B)}{\chi_2^S(T,\hmu_B)} \right) 
    \label{us-exact}\\
    &\simeq& -\frac{1}{2}\Bigg( 1+3 \frac{\chi_{11}^{BS}(T,\hmu_B)}{\chi_2^S(T,\hmu_B)} \Bigg)\; .
    \label{us2}
\end{eqnarray}
We show QCD results for the correlation
between light and strange quark densities
on the pseudo-critical line in 
Fig.~\ref{fig:BS-QS-mu_pseudo}.
We note that the magnitude of the
light and strange quark correlations 
decrease with increasing $\hmu_B$ in
HRG model calculations as well as in the QCD calculation. This arises from
the increasing importance of multiple strange baryon contributions on the pseudo-critical line.

\section{Second order cumulants on the pseudo-critical line}

\subsection{Pseudo-critical line and experimentally determined freeze-out conditions}

\begin{figure}[t]
     \includegraphics[width=0.45\textwidth]{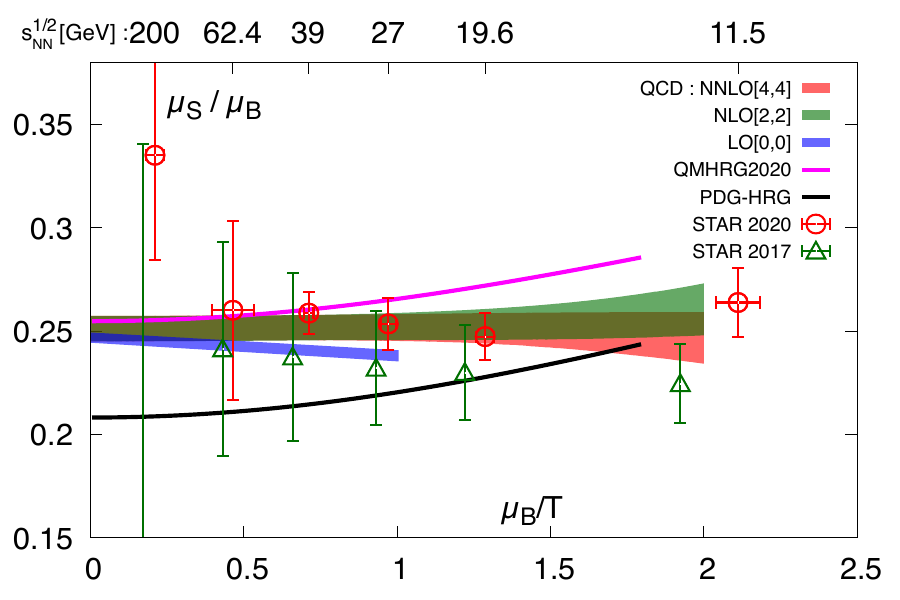}
    \caption{The ratio $\mu_S/\mu_B$ on the pseudo-critical line in (2+1)-flavor QCD versus $\hmu_B$ obtained for strangeness neutral matter with $n_Q/n_B=0.4$.  
The bands show continuum estimates for second and fourth 
order Taylor expansion results of 
$\mu_S/\mu_B$ on the pseudo-critical line 
that have been obtained from fits 
to data obtained in simulations
on lattices with temporal extent $N_\tau=8$ and 12.
    Also shown are results obtained by the STAR Collaboration \cite{STAR:2010yyv,Adamczyk:2017iwn,STAR:2019bjj} and HRG model calculations using
    the PDG-HRG and QMHRG2020 particle
    lists, respectively.}
    \label{fig:muBmuS-QCD}
\end{figure}

On the pseudo-critical line, $T_{pc}(\hmu_B)$, the ratio of
strangeness and baryon chemical potentials
obtained by demanding $n_S=0$ varies little
with increasing $\hmu_B$. Taylor expansion 
results for $\mu_S/\mu_B$ are shown in Fig.~\ref{fig:muBmuS-QCD} and are compared
with experimental results obtained by the STAR Collaboration by either fitting
particle yields for a large set of strange and non-strange hadrons to a HRG motivated hadronization model \cite{Adamczyk:2017iwn} or by fitting only yields of strange
baryons \cite{Adamczyk:2017iwn,STAR:2019bjj}.
Also shown in this figure are values for 
$\mu_S/\mu_B$ obtained in HRG model calculations based on the PDG-HRG and 
QMHRG2020 lists for hadron resonances. 
It is apparent that the former does not
describe the QCD results well, whereas
the QMHRG2020 provides a good description
of $\mu_S/\mu_B$ on the pseudo-critical
line at least for $\hmu_B\lsim 1$. Beyond this small deviations from the QCD results also show up. 
HRG model results based 
on PDG-HRG and QMHRG2020 differ by about 
20\%. This is similar to the differences found for the ratio
$\chi_{11}^{BS}/\chi_2^S$.
The good agreement between QCD 
and HRG model calculations based on QMHRG2020 at $T_{pc}(\hmu_B)$ in 
calculations of $\chi_{11}^{BS}/\chi_2^S$,
along with the agreement between determinations of the experimentally determined 
$\mu_S/\mu_B$ with QCD results, thus
suggests that the experimental results
for $\mu_S/\mu_B$ are indeed
sensitive to strange 
particle content at the time of freeze-out.

We start our discussion of cumulant 
ratios at finite temperature and non-zero values of the chemical potentials by
revisiting the analysis of 
the ratio of mean, $\chi_1^B(T,\hmu_B)$, and variance,
$\chi_2^B(T,\hmu_B)$, of net baryon-number density distributions

\begin{eqnarray}
R_{12}^B &=& \frac{\chi_1^B(T,\hmu_B)}{\chi_2^B(T,\hmu_B)}
\nonumber \\
&=&\hmu_B \bigg(1+ s_1\frac{ \chi_{11}^{BS}}{\chi_2^B}+q_1\frac{\chi_{11}^{QS}}{\chi_2^B}\bigg) +{\cal O}(\hmu_B^3)
\; ,
\label{R12b}
\end{eqnarray}
where we again choose 
 $n_S=0$, and $n_Q/n_B=0.4$. 

\begin{figure}[t]
    \includegraphics[width=0.49\textwidth]{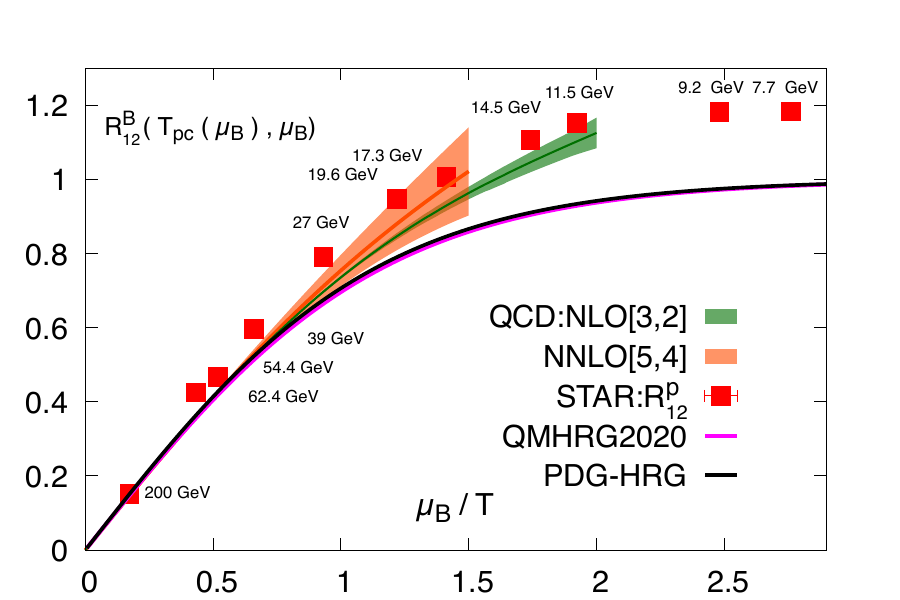}
    \caption{The ratio $R_{12}^B(T(\mu_B),\mu_B)$ on the pseudo-critical line in (2+1)-flavor QCD versus $\hmu_B$ obtained for strangeness neutral matter with $n_Q/n_B=0.4$. 
    The bands show continuum estimates for next-to-leading
    (NLO [3,2]) and next-to-next-to-leading (NNLO [5,4])
order Taylor expansion results 
on the pseudo-critical line 
that have been obtained from fits 
to data obtained in simulations
on lattices with temporal extent $N_\tau=8$ and 12 (see also discussion in the text).
Also shown are results obtained by the STAR Collaboration for the corresponding proton cumulant ratio, $R_{12}^p$ for various beam energies 
$\sqrt{s_{_{NN}}}$\cite{STAR:2021iop}.
Results from \cite{Adamczyk:2017iwn} have
been used to convert
    $\sqrt{s_{_{NN}}}$ to thermal parameters $(T_f,\hmu_B^f)$. Furthermore, we show
    recent results for
    $R_{12}^p$ obtained by the STAR Collaboration
    \cite{STAR-CPOD-C2C1}. For 
    data taken at two new beam energies,
    $\sqrt{s_{_{NN}}}=9.2$~GeV and 17.3~GeV we used the parametrization
    of interpolating curves for
    freeze-out parameters given in \cite{Andronic:2017pug}. Solid curves show
    results from HRG model calculations using
    the PDG-HRG and QMHRG2020 particle
    lists, respectively.}
    \label{fig:R12B}
\end{figure}

The ratio  $R_{12}^B(T,\hmu_B)$ has been analyzed by us previously
and was shown in \cite{Bazavov:2020bjn} in 
a $\hmu_B$-range relevant for the analysis of higher order cumulants. 
In Fig.~\ref{fig:R12B}
we show  $R_{12}^B(T,\hmu_B)$ in a 
larger range 
of $\hmu_B$ values. 
The continuum estimates based on
up to ${\cal O}(\hmu_B^5)$ Taylor 
series for the net baryon-number density,
$\chi_1^B$, and a ${\cal O}(\hmu_B^4)$ Taylor series for the net baryon number fluctuations, $\chi_2^B$
(NNLO [5,4]) shown in this figure are
statistically well controlled for $\hmu_B\le 1.5$. For large values of 
the chemical potential, $\hmu_B\le 2$,
we only show the result (NLO [3,2]) based on
${\cal O}(\hmu_B^3)$ and ${\cal O}(\hmu_B^2)$ Taylor series for 
$\chi_1^B$ and $\chi_2^B$, respectively. 
These continuum estimates are obtained
from fits
to data taken on lattices with temporal extent $N_\tau=8$ and $12$.
As can be seen, the ratio $R_{12}^B$ starts
to deviate from HRG model calculations at
about $\hmu_B\simeq 1$. The Taylor series, 
however, is still well controlled at least 
up to $\hmu_B\simeq 1.5$. 
We compare $R_{12}^B(T,\hmu_B)$, calculated
in (2+1)-flavor QCD, with the corresponding
ratio of net proton-number and its variance,
$R_{12}^p(\sqrt{s_{_{NN}}})$, \cite{STAR:2021iop} as follows. In order to convert
the experimental control parameter, $\sqrt{s_{_{NN}}}$, to a chemical potential
value on the freeze-out line we use the 
set of freeze-out temperatures ($T_{f}$)
and baryon chemical potentials ($\mu_B^f$)
determined from the analysis of particle
yields using fits based on the thermodynamics of a hadron gas in the Grand Canonical
Ensemble \cite{Adamczyk:2017iwn,STAR:2021iop}.

As can be seen in the figure, the 
experimental results for $R_{12}^p(T_f,\mu_B^f)$ are close to
the line of values for  $R_{12}^B(T,\mu_B)$ on the pseudo-critical line. 
We thus determined a set of 
freeze-out parameters obtained by
demanding $R_{12}^p(\sqrt{s_{_{NN}}})=
R_{12}^B(T_{pc}(\mu_B^f),\mu_B^f)$.
In Table~\ref{tab:freeze-out} we
compare the thus determined set of freeze-out
parameters, $\{ \mu_B^f,T_{pc}(\mu_B^f)\}$, with the experimental set of
freeze-out parameters, $\{ \mu_B^f,T_{ch}\}$, obtained by
comparing measured particle yields
to hadronization models
\cite{Andronic:2017pug,Adamczyk:2017iwn,ALICE:2023ulv}. 

As can be seen in Fig.~\ref{fig:R12B}, NLO and NNLO QCD results agree well with each other up to $\hmu_B\simeq 1.5$, where the errors of the NNLO expansion start getting large. 
In Tab.~\ref{tab:freeze-out}, we tabulate the
freeze-out parameters determined 
from $R_{12}^p$ using NLO QCD results for
$R_{12}^B(T,\hmu_B)$ and they agree well with
the freeze-out parameters $\{ T_f,\hmu_B^f\}$ obtained from particle yields down to 
$\sqrt{s_{_{NN}}}\simeq 17.3$~GeV, which also corresponds to $\hmu_B\simeq 1.5$.
For smaller $\sqrt{s_{_{NN}}}$ or
$\hmu_B\simeq 2.0$ the NNLO lattice QCD
results have too large errors for a 
detailed quantitative comparison.
Results from the NLO expansion are,
however, still in good agreement with 
the STAR data down
to $\sqrt{s_{_{NN}}}\simeq 11.5$~GeV.
In order to compare QCD results with STAR data
at even lower beam energies statistically well controlled
higher order Taylor series will be 
necessary.

It also should be noted that
the experimentally determined $R_{12}^p$ 
becomes larger than unity for $\sqrt{s_{_{NN}}}\lsim 17.3$~GeV. This is consistent with lattice QCD results for 
$R_{12}^B$, which become larger than unity
for $\hmu_B\gsim 1.3$ or $\mu_B\gsim 200$~MeV.
On the other hand, HRG model calculations 
based on non-interacting, point-like hadrons
will always lead to $R_{12}^B(T,\hmu_B)<1$. The experimental
data thus seem to reflect interactions
in strong-interaction matter at freeze-out (on the pseudo-critical line)
that go beyond those taken care of
in HRG models through the presence of
a tower of excited states and resonances.

We give a comparison
of the chemical potentials and pseudo-critical temperatures on the pseudo-critical line, 
$\{ T_{pc}(\hmu_B),\mu_B\}$, that are obtained by demanding $R_{12}^B=R_{12}^p$
and the freeze-out 
parameters, $\{ T_f,\mu_B^f\}$,
obtained from HRG model fits to 
measured hadron yields in 
Table~\ref{tab:freeze-out}

\begin{table*}[tb]
\begin{center}
\begin{tabular}{|c||rrrr||rrrr|}
\hline
\multicolumn{1}{|c||}{RHIC beam energy} & \multicolumn{4}{c||}{QCD, pseudo-critical line: using $R_{12}^p$} & \multicolumn{4}{c||}{HIC, freeze-out line: using yields} \\
\hline
$\sqrt{s_{_{NN}}}$ [GeV] & $\mu_B$ [MeV] & $\mu_S$ [MeV] & $T_{pc}$ [MeV] & ~ & $\mu_B$ [MeV] & $\mu_S$ [MeV] & $T_{ch}$ [MeV] & ~ \\
\hline
7.7 & --- & --- & --- & ~ & 398.2(16.4) & 89.5(6.0) & 144.3(4.8) & ~ \\
9.2 & --- & --- & --- & ~ & 358.3 & --- & 144.4 & ~ \\
11.5 & 297.9(12.4) & 73.5(4.8) & 148.9(2.9) & ~ & 287.3(12.5) & 64.5(4.7) & 149.4(5.2) & ~ \\
14.5 & 274.8(9.1) & 68.2(3.7) & 150.1(2.6) & ~ & 264 & --- & 151.6 & ~ \\
17.3 & 245.3(6.5) & 61.3(2.8) & 151.6(2.2) & ~ & 218.6 & --- & 154.7 & ~ \\
19.6 & 222.3(4.7) & 55.7(2.2) & 152.5(2.0) & ~ & 187.9(8.6) & 43.2(3.8) & 153.9(5.2) & ~ \\
27.0 & 170.5(2.1) & 42.9(1.3) & 154.2(1.6) & ~ & 144.4(7.2) & 33.5(3.6) & 155.0(5.1) & ~ \\
39.0 & 120.2(1.3) & 30.3(1.2) & 155.4(1.5) & ~ & 103.2(7.4) & 24.5(3.8) & 156.4(5.7) & ~ \\
54.4 & 89.3(0.9) & 22.5(0.6) & 155.9(1.5) & ~ & 83 & --- & 160.0 & ~ \\
62.4 & 81.5(0.8) & 20.5(0.5) & 156.0(1.5) & ~ & 69.2(5.6) & 16.7(3.3) & 160.3(4.9) & ~ \\
200 & 28.3(0.3) & 7.1(0.2) & 156.4(1.5) & ~ & 28.4(5.8) & 5.6(3.9) & 164.3(5.3) & ~ \\
\hline
\end{tabular}
\end{center}
\caption{The left hand column lists the various RHIC beam energies, $\sqrt{s_{_{NN}}}$, used by the STAR collaboration \cite{STAR:2021iop,STAR-CPOD-C2C1}
for their measurements of net proton-number cumulants and $B-S$ correlations. 
The three columns in the middle give the chemical potentials and the temperature values on the pseudo-critical line that correspond to $R_{12}^B=R_{12}^p$. The last three columns show corresponding freeze-out parameters obtained by fitting particle yields to a hadronization model based on the grand canonical ensemble \cite{Adamczyk:2017iwn}. Results for $\sqrt{s_{_{NN}}}=14.5$~GeV and 54.4~GeV are taken from \cite{STAR:2021iop}.
For $\sqrt{s_{_{NN}}}=9.2$~GeV and 17.3~GeV we used the parametrization
of interpolating curves for
freeze-out parameters given in \cite{Andronic:2017pug}.
}
\label{tab:freeze-out}
\end{table*}

\subsection{Strangeness correlations on the
pseudo-critical line}

\begin{figure*}[t]
\includegraphics[width=0.45\textwidth]{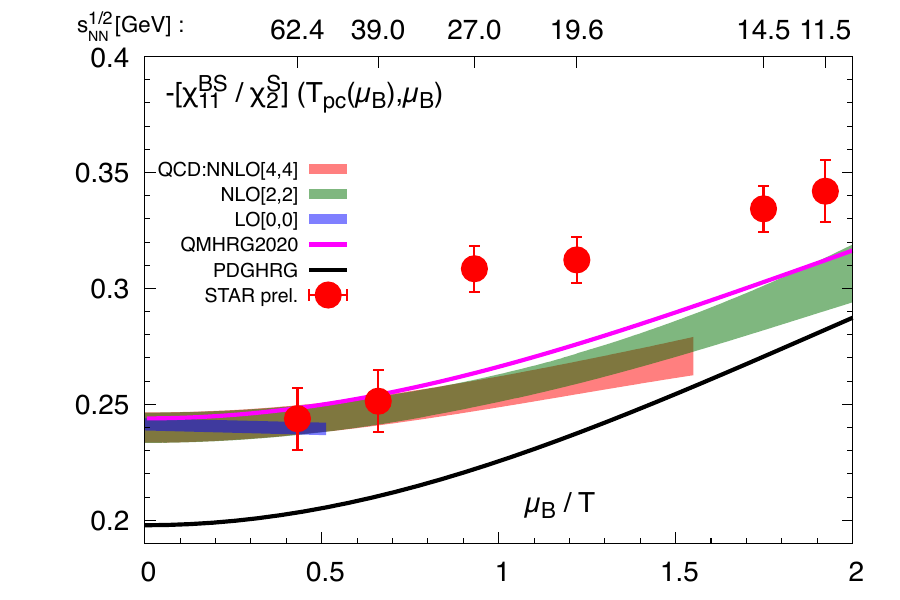}
\includegraphics[width=0.45\textwidth]{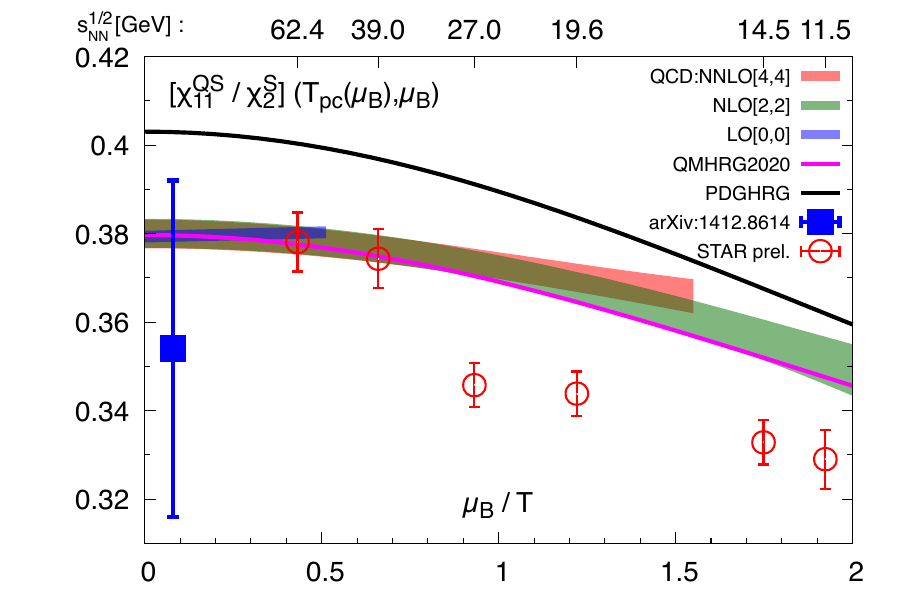}
\caption{{\it Left:} Correlation between net strangeness and net baryon-number densities normalized to the second order cumulant of strangeness fluctuations on the pseudo-critical line.
The bands show continuum estimates for second and fourth 
order Taylor expansion results
on the pseudo-critical line 
that have been obtained from fits 
to data obtained in simulations
on lattices with temporal extent $N_\tau=8$ and 12. In the case of 
fourth order expansion we show results only up to $\mu_B/T=1.5$.
{\it Right:} same as left hand figure
but for correlations between net strangeness and net electric charge densities. The data
shown in the left hand figure use the preliminary results for $\chi_{11}^{BS}/\chi_2^S$ obtained by the STAR
Collaboration \cite{STAR-CPOD}. Data points shown in the right hand
figure are obtained from the STAR data
making use of the relation given in
Eq.~\ref{BS-QS-rel}. Also shown in this figure is the result obtained in \cite{Braun-Munzinger:2014lba} from an analysis of ALICE data on strange particle yields (see also discussions in the text). 
}
\label{fig:BSmu_pseudo}
\end{figure*}

In Fig.~\ref{fig:BSmu_pseudo}~(left) we show results for the ratio $\chi_{11}^{BS}/\chi_2^S$ 
and compare with preliminary results for this
ratio obtained by the STAR Collaboration
\cite{STAR-CPOD}. As
can be seen the Taylor series for the ratio $\chi_{11}^{BS}/\chi_2^S$ converges well at least for $\hmu_B\lsim 1.5$. Similar to the observation made for other observables results in (2+1)-flavor QCD and HRG model calculations
based on the QMHRG2020 particle list also agree
well in this $\hmu_B$-range. The 
preliminary STAR results, however, agree well with the
QCD result only for $\sqrt{s_{_{NN}}}=62.4$~GeV 
and 39~GeV. For smaller $\sqrt{s_{_{NN}}}$ differences are
significant.
We note that this difference does not arise from a possible 
ambiguity in relating values of $\sqrt{s_{_{NN}}}$ to values for $\hmu_B$ derived
from experimental data.

As discussed above correlations between electric charge and strangeness are closely
related to $\chi_{11}^{BS}$. On the 
pseudo-critical line and for conditions met
in heavy ion collisions ($n_S=0, n_Q/n_B=0.4$)
violations of the sum rule, Eq.~\ref{BS-QS-sumrule}, are smaller than 0.5\% for $\hmu_B\le 2$, which covers 
almost the entire range of the RHIC BES-II energy range in collider mode.
In Fig.~\ref{fig:BSmu_pseudo}~(right) we 
show the ratio $\chi_{11}^{QS}/\chi_2^S$
on the pseudo-critical line and also
compare with preliminary STAR results for $\chi_{11}^{BS}/\chi_2^S$ \cite{STAR-CPOD} that have been
converted to $\chi_{11}^{QS}/\chi_2^S$
using Eq.~\ref{BS-QS-sumrule}. 

Also shown
in this figure is a result for $\chi_{11}^{QS}/\chi_2^S$ obtained in
\cite{Braun-Munzinger:2014lba} from 
particle yields measured by the ALICE
collaboration. This result has been obtained
using data on strange particle yields, obtained by the ALICE Collaboration at the LHC \cite{ALICE:2013mez,ALICE:2014jbq}.
The second order cumulants
$\chi_2^S,\ \chi_{11}^{BS}$ and $\chi_{11}^{QS}$ have been constructed from
measured particle yields taking into account
feed down corrections from $\phi$-mesons
and neutral kaons \cite{Braun-Munzinger:2014lba}. The second 
order cumulants, constructed in this way, obey
the QCD sum rule, Eq.~\ref{BS-QS-sumrule},
except for contributions arising from the 
feed down corrections. As pointed out in
\cite{Braun-Munzinger:2014lba} certain
decay channels, contributing in particular to
$\chi_{11}^{BS}$, are not known experimentally. Moreover, it is difficult to take care of
feed-down corrections arising from
decays of experimentally not well controlled higher kaon resonances 
and additional strange hadrons. This 
suggests that an accurate experimental
determination of $\chi_{11}^{BS}$ will
remain to be difficult without achieving better control over contributions arising from additional resonances and their decay channels.

\section{Conclusions}

We pointed out that the determination
of freeze-out parameters $( T_f,\mu_B^f, \mu_S^f)$ from particle
yields in heavy ion collisions is consistent with QCD parameters
describing thermal conditions realized on
the pseudo-critical line for strangeness
neutral matter with $n_Q/n_B=0.4$.
In particular, we find that at ($T_f,\mu_B^f, \mu_S^f)$, corresponding to $\sqrt{s_{_{NN}}}\ge 11.5$~GeV, the new results 
of the STAR collaboration obtained for the ratio of
mean net proton-number and the net proton number fluctuations, $R_{12}^p$, are
in good agreement with $R_{12}^B$ calculated in lattice QCD
on the pseudo-critical line. Similarly
we find that the ratio of strangeness and baryon chemical potentials, obtained from
strange baryon yields and QCD calculations,
respectively, are in good agreement.
This suggests that the values of 
thermal control parameters $(T,\mu_B,\mu_S)$, characterizing
strong-interaction matter created at
a given $\sqrt{s_{_{NN}}}$, are well
understood at least for $\sqrt{s_{_{NN}}}\ge 11.5$~GeV.

Furthermore, we presented QCD results for the
normalized baryon-number strangeness 
correlation $\chi_{11}^{BS}/\chi_2^S$
as function of the baryon chemical potential $\hmu_B$ on the pseudo-critical
line $T_{pc}(\hmu_B)$. The results,
based on fourth order Taylor series
of second order cumulants, are shown to
be well controlled for $\hmu_B\le 1.5$
or equivalently $\mu_B\lsim 200$~MeV.
For conditions realized in heavy ion collisions at the time of freeze-out this corresponds to beam energies $\sqrt{s_{_{NN}}}\ge 17.3$~GeV. 
Although results for $\chi_{11}^{BS}/\chi_2^S$ obtained in
QCD calculations agree well with 
experimental data for $\sqrt{s_{_{NN}}}= 62.4$~GeV and 39~GeV, we observe
significant differences for smaller
beam energies. We thus conclude that 
there is some tension between current experimental results on
strangeness and baryon number correlations and those appearing in
strong-interaction matter on the pseudo-critical line as described by equilibrium QCD thermodynamics. 
To verify or falsify that QCD
thermodynamics can provide a good 
description of second order cumulant 
ratios of fluctuations and correlations of conserved charges, as measured 
in heavy ion collisions using various
proxies, the origin of these differences clearly needs to be analyzed further. All data presented in the figures of this paper can be found in~\cite{epubdata:2992312}.

\vspace{0.5cm}
\section*{Acknowledgments}
This work was supported by the Deutsche Forschungsgemeinschaft
(DFG, German Research Foundation) Proj. No. 315477589-TRR 211; 
and the PUNCH4NFDI consortium
supported by the Deutsche Forschungsgemeinschaft (DFG, German Research Foundation) with project number 460248186 (PUNCH4NFDI).
This material is based upon work supported by The U.S. Department of Energy, Office of Science, Office of Nuclear Physics through Contract Nos.~DE-SC0012704, and within the frameworks of Scientific Discovery through Advanced Computing (SciDAC) award Fundamental Nuclear Physics at the Exascale and Beyond, and the NSFC under grant No.~12325508 and the National Key Research and Development Program of China under Grant No. 2022YFA1604900.

This research used awards of computer time  provided by the U.S. Department of Energy’s INCITE and ALCC programs at the Argonne and the Oak Ridge Leadership Computing Facilities. The Argonne Leadership Computing Facility at Argonne National Laboratory is supported by the Office of Science of the U.S. DOE under Contract No. DE-AC02-06CH11357. 
The Oak Ridge Leadership Computing Facility at the Oak Ridge National Laboratory is supported by the Office of Science of the U.S. DOE under Contract No. DE-AC05-00OR22725. 
Computations for this work were carried out in part on facilities of the USQCD Collaboration, funded by the Office of Science of the U.S. Department of Energy.

\appendix
\section{Expansions for second order cumulants
up to ${\cal O}(\hmu_B^4)$}
In Eqs.~\ref{BSNLO}-\ref{BNLO} we gave the ${\cal O}(\hmu_B^2)$
expansions for the second order cumulants $\chi_{11}^{BS}$, $\chi_{11}^{QS}$, $\chi_{2}^{S}$
and $\chi_{2}^{B}$. 
Here we give explicit expressions for expansions of these second order cumulants up to ${\cal O}(\hmu_B^4)$. This also requires expansions
of the strangeness and electric charge chemical 
potentials up to ${\cal O}(\hmu_B^3)$ as defined in 
Eqs.~\ref{muSexp}and \ref{muQexp}.
The coefficients $s_i$, $q_i$ appearing in these 
expansions are given in \cite{Bazavov:2017dus}.

With this we obtain for the second order cumulants
the expansions,
\begin{widetext}
\begin{eqnarray}
\chi_{11}^{BS}(T,\vec{\mu}) &=& \chi_{11}^{BS} + \frac{\hmu_B^2}{2!} [\chi_{31}^{BS} + 2s_1 \chi_{22}^{BS} + s_1^2 \chi_{13}^{BS} + 2 q_1 \chi_{211}^{BQS} + 2 q_1 s_1 \chi_{112}^{BQS} + q_1^2 \chi_{121}^{BQS}] \nonumber \\
&& + \frac{\hmu_B^4}{4!}  [ 24 \chi_{13}^{BS} s_1 s_3 + \chi_{15}^{BS} s_1^4 + 24 \chi_{112}^{BQS} q_1 s_3 + 24 \chi_{112}^{BQS} q_3 s_1  + 4 \chi_{114}^{BQS} q_1 s_1^3 + 24 \chi_{121}^{BQS} q_1 q_3 + 6 \chi_{123}^{BQS} q_1^2 s_1^2 
\nonumber \\
&& + 4 \chi_{132}^{BQS} q_1^3 s_1 + \chi_{141}^{BQS} q_1^4 + 24 \chi_{22}^{BS} s_3 + 4 \chi_{24}^{BS} s_1^3 + 24 \chi_{211}^{BQS} q_3 + 12 \chi_{213}^{BQS} q_1 s_1^2 + 12 \chi_{222}^{BQS} q_1^2 s_1\nonumber \\
&&   + 4 \chi_{231}^{BQS} q_1^3 
+ 6 \chi_{33}^{BS} s_1^2 + 12 \chi_{312}^{BQS} q_1 s_1 + 6 \chi_{321}^{BQS} q_1^2 + 4 \chi_{42}^{BS} s_1
+ 4 \chi_{411}^{BQS} q_1 + \chi_{51}^{BS}]+ {\cal O}(\hmu_B^6),  \\
\chi_{11}^{QS}(T,\vec{\mu})&=&\chi_{11}^{QS}+\frac{\hmu_B^2}{2!}\big[\chi_{211}^{BQS} + 2s_1\chi_{112}^{BQS} + s_1^2\chi_{13}^{QS} +  2 q_1\chi_{121}^{BQS} + 2 q_1s_1 \chi_{22}^{QS} + q_1^2 \chi_{31}^{QS}\big] \nonumber \\
&& + \frac{\hmu_B^4}{4!}[24 \chi_{013}^{BQS} s_1 s_3 + \chi_{15}^{QS} s_1^4 + 24 \chi_{022}^{BQS} q_1 s_3 + 24 \chi_{22}^{QS} q_3 s_1 + 4 \chi_{24}^{QS} q_1 s_1^3 + 24 \chi_{31}^{QS} q_1 q_3 + 6 \chi_{33}^{QS} q_1^2 s_1^2 \nonumber \\
&&+4 \chi_{42}^{QS} q_1^3 s_1 + \chi_{51}^{QS} q_1^4 
+ 24 \chi_{112}^{BQS} s_3 + 4 \chi_{114}^{BQS} s_1^3 + 24 \chi_{121}^{BQS} q_3 + 12 \chi_{123}^{BQS} q_1 s_1^2 + 12 \chi_{132}^{BQS} q_1^2 s_1  \nonumber \\
&&+ 4 \chi_{141}^{BQS} q_1^3 + 6 \chi_{213}^{BQS} s_1^2
+ 12 \chi_{222}^{BQS} q_1 s_1 + 6 \chi_{231}^{BQS} q_1^2 + 4 \chi_{312}^{BQS} s_1 + 4 \chi_{321}^{BQS} q_1
+ \chi_{411}^{BQS}] + {\cal O}(\hmu_B^6)\; , \\
\chi_{2}^{B}(T,\vec{\mu})&=& \chi_{2}^B +\frac{\hmu_B^2}{2!} \big[ \chi_4^B+s_1^2 \chi_{22}^{BS}+2 s_1\chi_{31}^{BS} + 2 q_1\chi_{31}^{BQ}+2 q_1 s_1 \chi_{211}^{BQS}+q_1^2\chi_{22}^{BQ} \big] \nonumber \\ 
&&+\frac{\hmu_B^4}{4!}[24 \chi_{22}^{BS} s_1 s_3 + \chi_{24}^{BS} s_1^4 + 24 \chi_{211}^{BQS} q_1 s_3
+ 24  \chi_{211}^{BQS} q_3 s_1 + 4 \chi_{213}^{BQS} q_1 s_1^3 + 24 \chi_{22}^{BQ} q_1 q_3 + 6 \chi_{222}^{BQS} q_1^2 s_1^2 \nonumber \\
&&+ 4 \chi_{231}^{BQS} q_1^3 s_1 + \chi_{24}^{BQ} q_1^4  + 24 \chi_{31}^{BS} s_3 + 4 \chi_{33}^{BS} s_1^3 + 24 \chi_{31}^{BQ} q_3 + 12 \chi_{312}^{BQS} q_1 s_1^2 + 12 \chi_{321}^{BQS} q_1^2 s_1 \nonumber \\
&& + 4 \chi_{33}^{BQ} q_1^3 + 6 \chi_{42}^{BS} s_1^2 + 12 \chi_{411}^{BQS} q_1 s_1 + 6 \chi_{42}^{BQ} q_1^2 
+ 4 \chi_{51}^{BS} s_1 + 4 \chi_{51}^{BQ} q_1 +  \chi_{6}^{B}]+{\cal O}(\hmu_B^6)  \\
\chi_{2}^{S}(T,\vec{\mu})&=& \chi_{2}^S +\frac{\hmu_B^2}{2!} [\chi_{22}^{BS} + s_1^2\chi_{4}^S + 2 s_1\chi_{13}^{BS} + 2q_1\chi_{112}^{BQS}+2q_1s_1\chi_{13}^{QS}+ q_1^2\chi_{22}^{QS}] \nonumber \\
&& +\frac{\hmu_B^4}{4!}[24\chi_{4}^{S}s_1s_3 + \chi_{6}^{S}s_1^4 + 24\chi_{13}^{QS}q_1s_3 + 24\chi_{13}^{QS}q_3s_1
+ 4\chi_{15}^{QS}q_1s_1^3 + 24\chi_{22}^{QS}q_1q_3 + 6\chi_{24}^{QS}q_1^2s_1^2  \nonumber \\
&&+ 4\chi_{33}^{QS}q_1^3s_1+ \chi_{42}^{QS}q_1^4 + 24\chi_{13}^{BS}s_3 + 4\chi_{15}^{BS}s_1^3 + 24\chi_{112}^{BQS}q_3 + 12\chi_{114}^{BQS}q_1s_1^2+ 12\chi_{123}^{BQS}q_1^2s_1 \nonumber \\ 
&& + 4\chi_{132}^{BQS}q_1^3 + 6\chi_{24}^{BS}s_1^2 + 12\chi_{213}^{BQS}q_1s_1 
+ 6\chi_{222}^{BQS}q_1^2 + 4\chi_{33}^{BS}s_1 + 4\chi_{312}^{BQS}q_1 + \chi_{42}^{BS}]+ {\cal O}(\hmu_B^6) \;
\end{eqnarray}
\end{widetext}

 \bibliography{bibliography}
\end{document}